\documentstyle[prd,aps,floats]{revtex}

\input cyracc.def
\font\tencyss=wncyss10
\def\cyss{\tencyss\cyracc}
\def\cb{\hbox{\cyss B}}

\begin{document}
\draft

\input epsf
\renewcommand{\topfraction}{1.0}
\twocolumn[\hsize\textwidth\columnwidth\hsize\csname 
@twocolumnfalse\endcsname

\preprint{Cologne-THP-H-99/03, gr-qc/9904067}

\title{Spacetime metric from linear electrodynamics}

\author{Yuri N. Obukhov \footnote{On leave from: Department of 
Theoretical Physics, Moscow State University, 117234 Moscow, 
Russia}\footnote{Electronic address: yo@thp.uni-koeln.de}\and
Friedrich W. Hehl \footnote{Electronic address: hehl@thp.uni-koeln.de}}
\address{Institute for Theoretical Physics, University of Cologne,
D--50923 K\"oln, Germany}


\maketitle

\medskip
\centerline{Received 4 May 1999; received in revised form 31 May 1999}

\centerline{Editor: R. Gatto}\medskip

\begin{abstract} The Maxwell equations are formulated on an arbitrary 
  $(1+3)$-dimensional manifold. Then, imposing a (constrained) linear
  constitutive relation between electromagnetic field $(E,B)$ and
  excitation $({\cal D},{\cal H})$, we {\em derive} the metric of
  spacetime therefrom. {\em file metric3.tex, 1999-05-30}
\end{abstract}

\vskip2pc]

In Einstein's gravitational theory, the {\em metric} $g$ of
spacetime is the fundamental field variable. The metric governs
temporal and spatial distances and angles. All other geometrical
notions can be derived from the metric. In particular, the affine
properties of spacetime, i.e.\ those related to parallel transfer and
represented by the linear {\em connection} $\Gamma$, are subordinate
to the metric. In Riemannian spacetime, the connection $\Gamma$ can be
exclusively expressed in terms of the metric, and the same is true for
the curvature $R\sim curl\, \Gamma$.

In {\em gauge}-theoretical approaches to gravity, the connection
$\Gamma$ becomes an independent field variable. In the viable
Einstein-Cartan theory of gravity, e.g., it is Lorentz algebra valued
and couples to the spin of matter. In fact, in such approaches, the
coframe $\vartheta$ and the connection $\Gamma$ appear as fundamental
variables. The metric is sort of absorbed by a priori choosing the
coframe to be pseudo-orthonormal and the connection to be Lorentz
algebra valued.

In any case, the dichotomy of metric and affine properties of
spacetime and attempts to understand it runs through much of
present-day theorizing on the fundamental structure of spacetime.
Already Eddington, e.g., tried to derive $g$ from (a symmetric)
$\Gamma$ by choosing a suitable gravitational Lagrangian $\sim |\det
{\rm Ric}_{ij}(\Gamma,\partial\Gamma)|^{1/2}$, see \cite{Schro}; here
and in future $i,j,k\dots=0,1,2,3$ are coordinate indices. Then
$\Gamma$ would be fundamental and $g$ be a derived concept.

Thinking more from the point of view of a quantum substructure
supposedly underlying classical spacetime, arguments were advanced
\cite{Sach} that the metric is some kind of effective field which
``froze out'' during a cosmic phase transition in the early universe, 
cf.\ also \cite{Fink}. In other words, also here the metric would be a
secondary structure comparable to the strain field in a solid.

The aim of this letter is, however, more modest: We formulate
classical electrodynamics in the so-called metric-free version, see
\cite{Truesdell,Post62,Stachel}, by taking recourse to the
conservation laws of electric charge and magnetic flux and to the
existence of a Lorentz force density. To complete the apparatus of the
field equations, we have eventually to specify the constitutive
relation between field strength $F=(E,B)$ and excitation 
$H=({\cal D},{\cal H})$.

We choose a {\em linear} law $H_{ij}=\kappa_{ij}{}^{kl}F_{kl}/2$ with
$21$ independent constitutive functions $\kappa_{ij}{}^{kl}(x)$. The
linear law can be interpreted as a new kind of duality operation $\#$
mapping 2-forms into 2-forms: $H\sim{}^\#F$. Let us stress that 
{\em no} metric has been used so far.

We impose a constraint on the duality operator, namely $^{\#\#}=-1$
(for Euclidean signature $+1$). This, together with two formulas of
Urbantke \cite{Urban} which had been used in a Yang-Mills context,
allows us to derive from $\kappa_{ij}{}^{kl}$ a metric with
pseudo-Euclidean signature. In this way we recognize of how closely
the concept of a metric is connected with the electromagnetic
properties of spacetime itself or of material media embedded therein
-- a fact which, perhaps, doesn't come as a surprise in view of the
principle of the constancy of the velocity of light.  \vskip2pt

\noindent(1) {\em Three axioms of electrodynamics:}  
Spacetime is assumed to be a 4-dimensional differentiable manifold
which allows a foliation into 3-dimensional submanifolds which can be
numbered with a monotonically increasing parameter $\sigma$. The
existence of an electric current (3-form) $J=-j\wedge d\sigma+\rho$ is
postulated which, by axiom 1, is conserved: 
\begin{equation}\label{axiom1} 
\oint\limits_{C_3}J=0\,,\quad
  \partial C_3=0\,.
\end{equation} Here $C_3$ is an arbitrary closed
3-dimensional submanifold of the 4-manifold.  By de Rham's theorem,
the inhomogeneous Maxwell equation is a consequence therefrom,
\begin{equation}
J=dH\,,
\end{equation} 
with $H=H_{ij}dx^i\wedge dx^j/2= -{\cal
  H}\wedge d\sigma+{\cal D}$.  The current $J$, together with a force
density $f_\alpha$, originating from mechanics, allow to formulate the
Lorentz force density as axiom 2:
\begin{equation}\label{axiom2}
  f_\alpha= (e_\alpha\rfloor F) \wedge J\,.
\end{equation} 
Greek indices $\alpha,\beta,\dots=0,1,2,3$ are
anholonomic or frame indices and $e_\alpha$ is the local frame, the
interior product is denoted by $\rfloor$. This axiom introduces the
electromagnetic field strength (2-form) $F=F_{ij}dx^i\wedge
dx^j/2=E\wedge d\sigma+B$ as a new concept. In axiom 3, the
corresponding magnetic flux is assumed to be conserved,
\begin{equation}\label{axiom3}
  \oint\limits_{C_2}F=0\,,\quad \partial C_2=0\,,
\end{equation} 
for any closed submanifold $C_2$. As a consequence, we find the
homogeneous Maxwell equation 
\begin{equation}
dF=0\,.
\end{equation}

These equations are all diffeomorphism invariant and don't depend on
metric or connection. This is also true for the exterior {\em and} the
interior product. Therefore these equations are valid in special and
general relativity likewise and in non-Riemannian spacetimes of gauge
theories of gravity, see \cite{Punt}. Electric charges and, under
favorable conditions, also magnetic flux quanta can be counted. This
is why the metric is dispensible under those circumstances.

\noindent(2) {\em Constitutive law as axiom 4:} The simplest constitutive 
law is, of course, a linear law. If it is additionally isotropic, it
yields in particular vacuum electrodynamics. However, isotropy can
only be formulated if a metric is available which is not the case
under the present state of our discussion. Therefore we assume only
linearity: 
\begin{equation}
H_{ij}={1 \over 2}\,\kappa_{ij}{}^{kl}(x)\,F_{kl}\,.\label{kappa}
\end{equation}
Since $H$ is an odd and $F$ an even form, the constitutive matrix $\kappa$ 
is odd. Therefore we split off the Levi-Civita symbol
\begin{equation}\label{chi}
  \tilde{\chi}_{ijkl}=\frac{1}{2}\,\kappa_{ij}{}^{mn}\,
  \epsilon_{mnkl}=f{\chi}_{ijkl}\,,
\end{equation}
where $f$ is a dimensionful scalar function such that $\chi_{ijkl}$ is
dimensionless. The tensor density $\chi_{ijkl}$ carries the weight
$-1$. The Lagrangian of the theory is quadratic in $F$. Thus we find
$\chi_{ijkl}=\chi_{klij}$, i.e., $21$ independent functions.
Since $H_{ij}$ and $F_{kl}$ can be measured independently, the 
constitutive functions $\chi_{ijkl}$ can be experimentally determined. 

It is convenient to write $H$ and $F$ as row vectors $H_I=(H_{01},
H_{02},H_{03},H_{23},H_{31},H_{12})$, where $I$ runs from $1$ to $6$,
etc. Then the constitutive law reads
\begin{equation}\label{chiHF1}
  H_I=\kappa_I{}^KF_K=\chi_{IM}\,\epsilon^{MK}fF_K\,\>{\rm
    with}\>\chi_{IM}=\chi_{MI}\,.
\end{equation}
Furthermore, in local coordinates, the basis of the 2-forms is
represented by the six 2-forms $dx^i\wedge dx^j$. They can be put into
the column vector Cyrillic $B$, namely $\cb^I$. Then $H=H_I\cb^I$ and
$F=F_I\cb^I$.

\noindent(3) {\em Duality operator $\#$ and its closure:} 
We can define, by means of the linear constitutive law (\ref{chiHF1}),
a new {\em duality operator} mapping 2-forms into 2-forms.
Accordingly, we require for the 2-form basis
\begin{equation}\label{sharp0}
{}^\#\cb ^I=\left(\chi_{KM}\,\epsilon^{MI}\right)\cb ^K\,,
\end{equation}
i.e., the duality operator incorporates the constitutive properties
specified in (\ref{chiHF1}). In particular, we have for the electromagnetic
field two-forms $H=f{}^\# F$.

A duality operator, applied twice, should, up to a sign, lead back to
the identity. By such a postulate we can constrain the number of
independent components of $\chi$ without using, say, a metric:
\begin{equation}\label{closure} 
{}^{\#\#}=-1\,.
\end{equation} 
We concentrate here on the minus sign; the rule $^{\#\#}= +1$ would
lead to Euclidean signature.

It is convenient to write the $6\times 6$ matrices, which define the 
duality operator, in terms of $3\times 3$-matrices:
\begin{equation}\label{xixi1}
  \chi_{IK}=\chi_{KI}= \left(\begin{array}{cr}A& C \\C^{{\rm T}} &B
    \end{array}\right)\,,\> \epsilon^{IK} =\epsilon^{KI}= 
  \left(\begin{array}{cr}0& {\mathbf 1} \\{\mathbf 1}
      &0\end{array}\right)
\end{equation}
Here $A=A^{\rm T}\,,B=B^{\rm T}$, the superscript ${}^{\rm T}$ denoting
transposition.  The general non-trivial solution of (\ref{closure})
is given by 
\begin{equation}\label{xixi3}
  \chi_{IK}= \left(\begin{array}{cc}pB^{-1} + qN& B^{-1}K \\ -KB^{-1}
      &B \end{array}\right)\,,
\end{equation}
Here $B$ is a nondegenerate arbitrary {\em symmetric} matrix (6 
independent components) and $K$ an arbitrary {\em antisymmetric} matrix 
(3 components). Furthermore, we construct the symmetric matrix $N$ as a 
solution of the homogeneous system $KN=NK=0$. With $K_{ab}=\epsilon_{abc}
k^c$, we have explicitly: 
\begin{equation}
N =  \left(\begin{array}{ccc}
(k^1)^2& k^1  k^2 & k^1  k^3 \\ k^1k^2&(k^2)^2& k^2  k^3 \\ 
 k^1  k^3  & k^2  k^3  &(k^3)^2\end{array}\right)\,.\label{Ndia}
\end{equation}
Finally $q = - 1/{\det B}$, $p = [{\rm tr}(NB)/\det B] - 1$.

\noindent (4) {\em Selfduality and a triplet of 2-forms:} With our new 
duality operator we can define the selfdual (s) and the anti-selfdual
(a) of a 2-form. For the 2-form basis $\cb$ we have
\begin{equation}
  {\stackrel {({\rm s})} \cb}:={\frac 1
    2}(\cb\,-i\,^\#\!\cb),\label{scb}\,,\quad {\stackrel {({\rm a})}
    \cb}:={\frac 1 2}(\cb\,+i\,^\#\!\cb)\,,\label{acb}
\end{equation}
with $^\#{\stackrel {({\rm s})} \cb}= i{\stackrel {({\rm s})} \cb}$,
$^\#{\stackrel {({\rm a})} \cb}= -i{\stackrel {({\rm a})} \cb}$. Thus
the 6-dimensional space of 2-forms decomposes into two 3-dimensional
invariant subspaces corresponding to the eigenvalues $\pm i$ of the
duality operator. With the decomposition into two 3-dimensional row
vectors,
\begin{equation}\label{cb3x2}
\cb^I = \left(\begin{array}{c} \beta^a \\ \gamma^b
\end{array}\right)\,,\quad a,b, \dots = 1,2,3\,,
\end{equation}
we take care of this fact also in the 2-form basis. 

One of the 3-dimensional invariant subspaces can be spanned by, say,
${\stackrel {({\rm s})} \gamma}$, whereas ${\stackrel {({\rm s})}
  \beta}$ is obtained from it by means of the linear transformation
${\stackrel {({\rm s})} \beta}=(i + B^{-1}K)B^{-1}{\stackrel {({\rm
      s})} \gamma}$. Therefore ${\stackrel {({\rm s})} \gamma}$
subsumes the properties of this invariant subspace and so does the
triplet of 2-forms
\begin{equation}\label{triplet}
S^{(a)}:= -(B^{-1})^{ab}\,{\stackrel{({\rm s})}{ \gamma}}{}^b\,.
\end{equation}
Hereafter, $(B^{-1})^{ab}$ and $B_{cd}$ denote the matrix elements of
$B^{-1}$ and $B$, respectively.  Incidentally, the anti-self dual 
${\stackrel {({\rm a})} \gamma}$ spans the other invariant subspace. The 
whole information of the linear constitutive law (\ref{chiHF1}) is now 
encoded into the triplet of 2-forms $S^{(a)}$. A direct calculation
demonstrates that they satisfy the so-called completeness condition:
\begin{equation}
S^{(a)}\wedge S^{(b)} = {\frac 1 3}\,
(B^{-1})^{ab}\,(B)_{cd}\,S^{(c)}\wedge S^{(d)}.\label{complSB}
\end{equation}

\noindent (5) {\em Extracting the metric:} Urbantke \cite{Urban} (see also 
the discussions in \cite{Harnett,Tert}) was able to derive, within
$SU(2)$ Yang-Mills theory, a 4-dimensional metric $g_{ij}$ (with $i,j,
\dots = 0,1,2,3$) from a triplet of 2-forms which are related to
2-plane elements of spacetime with certain distinguished properties.
Since the completeness condition (\ref{complSB}) is fulfilled, the
Urbantke's formulas
\begin{eqnarray}
  \sqrt{{\det}\,g}\,g_{ij} & =& -\,{\frac 2 3}\,\sqrt{\det
    B}\,\epsilon_{abc}\, \epsilon^{klmn}\,S^{(a)}_{ik}S^{(b)}_{lm}
  S^{(c)}_{nj}\,,\label{urbantke1}\\ \sqrt{{\det}\,g} & =& -\,{\frac 1
    6}\,\epsilon^{klmn}\,B_{cd}\,
  S^{(c)}_{kl}S^{(d)}_{mn}\,,\label{urbantke2}
\end{eqnarray}
are also applicable in our case. Here the $S^{(a)}_{ij}$ are the
components of the 2-form triplet according to $S^{(a)} = S^{(a)}_{ij}
dx^i\wedge dx^i/2$.

If we express ${\stackrel {({\rm s})} \gamma}$ in terms of $\beta$ and 
$\gamma$ and then substitute (\ref{triplet}) into (\ref{urbantke1}),
(\ref{urbantke2}), we can, after a very involved computation, display 
the metric explicitly:
\begin{equation}\label{gij}
g_{ij} = {\frac 1 {\sqrt{\det B}}}\left(\begin{array}{c|c}\det B &
-\,k_a \\ \hline -\,k_b & -\,B_{ab} + (\det B)^{-1}\,k_a\,k_b
\end{array}\right)\,.
\end{equation}
Here we used the abbreviation $k_a:=B_{ab}\,k^b = B_{ab}\,$ $ \epsilon
^{bcd}K_{cd}/2$. The $3\times 3$ matrix $B_{ab}$ can have any signature.
Nevertheless, Eq.(\ref{gij}) always yields a metric with Minkowskian
signature. 

This representation (\ref{gij}) of the metric is our basic result.
Since the triplet $S^{(a)}$ is defined up to an arbitrary scalar
factor, our procedure in general defines a {\em conformal} class of
metrics rather than a metric itself.

As the most simple example, we will construct the Minkowski metric.
Recall that (\ref{gij}) depends on the symmetric $3\times 3$ matrix
$B$ (not to be confused with the magnetic field $B$) and the
antisymmetric $3\times 3$ matrix $K$. If we choose $f^2 =
\varepsilon_0/\mu_0$, $B = (\varepsilon_0\mu_0)^{-1/2}\,{\mathbf 1}$
and $K = 0$, then, according to (\ref{chi}), this translates into the
conventional vacuum relations
\begin{equation}
\vec{\cal D}=\varepsilon_0\vec{E}\quad{\rm and}\quad
\vec{\cal H}=\vec{B}/\mu_0. \label{vacuum}
\end{equation}
On the other hand, if we substitute it into 
(\ref{gij}), we immediately find (denoting $c:=1/\sqrt{\varepsilon_0\mu_0}$)
\begin{equation}
g_{ij} = {\frac 1 {\sqrt{c}}}\left(\begin{array}{crrr}c^2 & 0 & 0 & 0 
\\ 0 & -1 & 0 & 0\\ 0& 0 & -1 & 0\\ 0& 0& 0& -1\end{array}\right),
\end{equation}
i.e.\ the Minkowski metric of special relativity (including its
signature) has been derived from the conventional vacuum relation
(\ref{vacuum}) between electromagnetic excitation $H=({\cal D},{\cal H})$ 
and field strength $F=(E,B)$. \vskip2pt

\noindent(6) {\em Outlook:} In this letter we have demonstrated that
the metric-free formulation of classical electrodynamics (in the
spirit of the old Kottler-Cartan-van Dantzig approach, cf.\
\cite{Post62}) naturally leads to the reconstruction of the spacetime
{\it metric from the constitutive law}. It seems worthwhile to remind
ourselves that the constitutive (or material) relation is a postulate
which arises not from a pure mathematical considerations but rather
from the analysis of experimental data \cite{Truesdell}. A development
of an alternative axiomatics of the Maxwell theory on the basis of the
postulate of the well-posedness of Cauchy problem \cite{Laemm} (see
also a related discussion in \cite{Stachel}) gives good reasons to
assume that the crucial closure condition (\ref{closure}) is
tantamount to a postulate of a well-posed Cauchy problem for Maxwell's
equations.

We are grateful to Claus L\"ammerzahl for helpful discussions on
Maxwell's theory and its axiomatic basis. This work was partially 
supported by the grant INTAS-96-842 of the European Union (Brussels). 

{\it Note added in proof:} After this work was completed, we have
learned that Sch\"onberg \cite{schoen}, in a not widely available
journal, had already developed the approach to the spacetime metric on
the basis of the constitutive relation (\ref{kappa}) and the
``closure'' relation (\ref{closure}). We are grateful to Helmuth
Urbantke and Ted Jacobson for corresponding remarks and to Jos\'e W.
Maluf for sending us a copy of Sch\"onberg's paper and his curriculum
vitae. However, our derivation of the general solution (\ref{xixi3})
of the ``closure" relation, and the complete explicit construction of
the metric (\ref{gij}) are new. Moreover, even earlier, Peres
\cite{peres} investigated related structures, and thus he can be
considered as a forerunner of Sch\"onberg. We thank Asher Peres for
pointing out to us the relevance of his paper. In the meantime
Guillermo Rubilar has checked our formula (\ref{xixi3}) by means of
computer algebra. For further discussions of our preprint we would
like to thank Ted Frankel, Yakov Itin, Bahram Mashhoon, and Eckehard
Mielke.

\end{document}